\documentclass[11pt]{article}

\usepackage[utf8]{inputenc}
\usepackage[T1]{fontenc}
\usepackage{geometry}
\usepackage{mathtools}  
\usepackage{amssymb}
\usepackage{amsthm}
\usepackage{graphicx}
\usepackage{mathrsfs}
\usepackage{xcolor}
\usepackage{booktabs}
\usepackage{caption}
\usepackage{enumitem}
\usepackage{braket}
\usepackage{comment}
\usepackage[hidelinks]{hyperref}
\usepackage{doi}
\usepackage{microtype}
\usepackage{cleveref}

\newcommand{\tr}{\operatorname{tr}}
\newcommand{\cumulantGeneratingFunction}{\psi}
\newcommand{\naturalParameters}{\boldsymbol{\theta}}
\newcommand{\naturalParameter}{\theta}

\newcommand{\gameTime}{\tau}
\newcommand{\entropyTime}{t}
\newcommand{\variables}{\mathbf{x}}

\newcommand{\localHamiltonian}{\left.\mathcal{H}_{\text{local}}\right.}

\newcommand{\sufficientStatisticExpectation}{\mu}
\newcommand{\sufficientStatistics}{\mathbf{T}}
\newcommand{\sufficientStatisticsExpectation}{\boldsymbol{\mu}}
\newcommand{\densityEigenvalue}{q}
\newcommand{\jointVonNeumannEntropy}{\mathsf{H}}
\newcommand{\marginalVonNeumannEntropy}{\mathsf{h}}
\newcommand{\jointShannonEntropy}{H}
\newcommand{\marginalShannonEntropy}{h}

\newif\ifdraft

\ifdraft
\newcommand{\tk}[1]{\textcolor{red}{\textbf{[Neil: #1]}}}
\else
\newcommand{\tk}[1]{}
\fi

\ifdraft
\usepackage{draftwatermark}
\SetWatermarkText{DRAFT}
\SetWatermarkScale{3}
\fi

\theoremstyle{definition}

\theoremstyle{remark}

\numberwithin{theorem}{section}
\numberwithin{proposition}{section}
\numberwithin{lemma}{section}
\numberwithin{corollary}{section}
\numberwithin{definition}{section}
\numberwithin{remark}{section}

\usepackage[authoryear,sort&compress,round]{natbib}
\ifdraft

\else
\excludecomment{draftcomment}
\fi

\begin{document}

\title{The Origin of the Inaccessible Game}

\author{Neil D. Lawrence\\
Computer Laboratory\\
University of Cambridge\\
15 J. J. Thomson Avenue\\
Cambridge CB3 0FD, UK\\
\url{ndl21@cam.ac.uk}}

\date{18th January 2026}

\maketitle

\begin{abstract}
  The inaccessible game is an information-geometric framework where
  the dynamics of information loss emerge from maximum entropy
  production under marginal-entropy conservation.

  We study the game's starting state, termed the origin. Classical
  Shannon entropy forbids a representation of the game's natural
  origin: non-negativity of conditional entropy rules out zero joint
  entropy with positive marginal entropies. Replacing Shannon entropy
  with von Neumann entropy within the Baez--Fritz--Leinster--Parzygnat
  (BFLP) categorical framework removes this obstruction and admits a
  well-defined origin: a globally pure state with maximally mixed
  marginals, selected up to local-unitary equivalence.

  At this locally maximally entangled (LME) origin, marginal-entropy
  conservation becomes a second-order geometric condition. Because the
  marginal-entropy sum is saturated termwise, the constraint gradient
  vanishes and first-order tangency is vacuous; admissible directions
  are instead selected by the kernel of the constraint Hessian, which
  is characterised by the marginal-preserving tangent space.

  We derive the constrained gradient flow in the matrix exponential
  family and show that, as the origin is approached, the game's affine
  time parameter degenerates. This motivates an axiomatically
  distinguished reparametrisation, \emph{entropy time} $\entropyTime$,
  defined by
  $\tfrac{\text{d}\jointVonNeumannEntropy}{\text{d}\entropyTime} = c$
  for a fixed constant $c>0$. In this parametrisation, the infinite
  affine-time approach to the boundary is mapped to a finite
  entropy-time interval.

  The constrained dynamics split into a symmetric dissipative
  component that realises steepest entropy ascent (SEA) and a
  reversible component that can be represented as unitary evolution.

  As in the classical game, marginal-entropy conservation is
  equivalent to conservation of a sum of local modular Hamiltonian
  expectations, a state-dependent ``modular energy''; in Gibbs regimes
  where the local modular generators become approximately
  parameter-invariant, this reduces to familiar fixed-energy
  constraints from nonequilibrium thermodynamics and a shared temporal
  representation emerges.
\end{abstract}

\section{Introduction}

The inaccessible game \citep{Lawrence-inaccessible25} is an
information geometric framework built on four axioms. The first three
Baez--Fritz--Leinster (BFL) axioms characterise information loss as
scaled entropy difference \citep{Baez-characterisation11}. The fourth,
information conservation, requires that marginal entropies remain
constant along any trajectory.

The dynamics of the game come from an information relaxation
principle: the system evolves by maximising joint entropy production
in natural parameter coordinates, $\nabla \jointShannonEntropy$, while
conserving the sum of marginal entropies, $\sum
\marginalShannonEntropy_i = C$.

This paper addresses a structural tension at the origin of the
game. The problem can be seen by rewriting the marginal entropy
constraint and introducing the multi-information
\citep{Watanabe-multiinformation60}, $I = \sum_i
\marginalShannonEntropy_i - \jointShannonEntropy$. This allows us to
rewrite the marginal entropy constraint as
$$
I + \jointShannonEntropy = C.
$$
The information relaxation dynamics suggest that the game begins
with maximal multi-information, i.e.\ $I = C$,
$\jointShannonEntropy=0$, and ends with maximal entropy, i.e.\ $I =
0$, $\jointShannonEntropy=C$. However, this initial condition cannot
be represented when Shannon's entropy underpins our notion of
information loss. The problem arises because the Shannon conditional
entropy is non-negative, $\jointShannonEntropy_{X|Y} \geq
0$. Fortunately, if we modify the axioms that define information loss
we can substitute the Shannon entropy with the von Neumann
\citep{vonNeumann-thermodynamik27,Wehrl-entropy78,Parzygnat-functorial22}.
The von Neumann entropy admits \emph{negative} conditional entropy
$\jointVonNeumannEntropy_{A|B}$ allowing the origin to be configured.

As in the classical game, we derive the dynamics of the modified game
using the exponential family. For the von Neumann entropy we require a
\emph{matrix exponential} form of the family. In this form sufficient
statistics are replaced by a Hermitian basis $\{F_a\}$ which is
globally sufficient. When we apply the information relaxation
principle from the classical game we see that the updated game
performs steepest entropy ascent on the curved matrix exponential
family.\footnote{The matrix exponential family is more commonly called
the quantum exponential family. We choose matrix exponential family
for our terminology as in our context the family is required for its
noncommutative properties rather than because of quantisation.} When
we \emph{linearise} the dynamics we recover a GENERIC structure where
the reversible sector can be represented in the Liouville/von Neumann
commutator form \citep{vonNeumann-book32}. We identify the
information-geometric structure induced by marginal-entropy
conservation and show that it selects a \emph{axiomatically
distinguished} (trajectory-wise) clock, entropy time. By
\emph{axiomatically distinguished} we mean: uniquely identifiable
within the game axioms, without introducing any additional external
choice such as Hamiltonians, spatial coordinates, a temperature scale,
or a background clock. We use this entropy time to parametrise the
constrained flow.

In the classical game, ``steepest entropy ascent'' referred to ascent
with respect to the \emph{Euclidean} inner product on
$\naturalParameters$-space, with the marginal-entropy constraint
enforced by Euclidean projection in these coordinates. To ensure that
the entire trajectory is axiomatically distinguished, in the modified
game we use the Riemannian metric given by the Fisher
information,\footnote{This Fisher metric is often called the
Bogoliubov--Kubo--Mori metric in the quantum literature.}  thereby
obtaining natural-gradient dynamics.

Our decomposition into reversible and irreversible components is not
itself novel, quantum GENERIC \citep[see e.g.][for a thermodynamic
  quantum master equation]{Ottinger-thermodynamic10} and closely
related metriplectic frameworks \citep{Turski-dissipative96} already
exist. Steepest entropy ascent has also been studied in quantum
thermodynamics \citep[see e.g.][for steepest entropy ascent quantum
  thermodynamics, SEAQT]{Beretta-quantum85,Beretta-fourth20}. Our
contribution is how these structures emerge from an axiomatic
foundation. In our derivation the conserved generator arises from the
marginal entropy constraint. This constraint aligns with the
Hamiltonian in Gibbsian regimes when the modular operators become
effectively parameter-invariant. The resulting constraints reduce to
the standard `fixed energy observable' constraints used in Jaynes's
maximum-entropy principle
\citep{Jaynes-information57a,Jaynes-information57b} and Beretta's
SEAQT.

\subsection{Contributions and Paper Roadmap}

The paper makes four main contributions.
\begin{itemize}
  \item We show that Shannon entropy structurally forbids the
    inaccessible game's natural origin through the non-negativity of
    conditional entropy.  We demonstrate that von Neumann entropy
    provides an axiomatic extension compatible with functorial
    information loss that admits a well-defined origin.

  \item We identify an axiomatically distinguished origin where the
    marginal-entropy conservation constraint collapses to a
    second-order condition: the constraint gradient vanishes, and
    admissibility is determined by the kernel of the constraint
    Hessian.

  \item We derive an axiomatically distinguished reparametrisation,
    entropy time, that uniformises entropy production and compactifies
    the infinite affine-time approach to the boundary into a finite
    interval, resolving the boundary degeneration of the natural
    parameter flow.

  \item We show that the resulting constrained dynamics admit a
    GENERIC-compatible (i.e.\ reversible--irreversible)
    decomposition. The conserved generators arise from marginal
    entropies via local modular Hamiltonians rather than from an
    assumed fixed energy observable. In effective Gibbs-locked regimes
    these may behave approximately as scaled fixed Hamiltonians.
\end{itemize}

The rest of the paper is organised as follows. In
Section~\ref{sec-origin-conflict} we recap the inaccessible game and
resolve the classical-quantum conflict at the origin. We show why
information isolation based on Shannon entropy forbids the origin and
how von Neumann entropy enables it. The shift to von Neumann entropy
encourages us to take an algebraic view of probability where
expectations form the primitives, not outcomes.
Section~\ref{sec-matrix-exponential} studies the flow in the matrix
exponential family and identifies the degeneracies at the
origin. Section~\ref{sec-entropy-time} introduces entropy time as the
trajectory-wise unique affine reparametrisation that uniformises
entropy production along the constrained flow. In
Section~\ref{sec-classical-vs-quantum-energy} we show how the marginal
entropy constraint reduces to a modular Hamiltonian in Gibbsian
regimes. When the regimes are approximate we can calibrate the entropy
time parameterisation with the quasi-reversible sector.

\section{Classical Conflict at the Origin}
\label{sec-origin-conflict}

The inaccessible game \citep{Lawrence-inaccessible25} depends on four
axioms and an information relaxation principle. The first three axioms
are taken from \citet{Baez-characterisation11}. They define
information loss of a process in terms of the difference between two
Shannon entropies, $\jointShannonEntropy$, measured before and after a
process. The fourth axiom is information conservation. It suggests
that the sum of marginal entropies should be conserved,
$$
\sum_i \marginalShannonEntropy = C,
$$
rendering the game `information-isolated'. 

An information relaxation principle drives the dynamics. By rewriting
the constraint in terms of the multi-information,
$$
I + \jointShannonEntropy = C,
$$
where $I = \sum_i \marginalShannonEntropy_i - \jointShannonEntropy$
we see that relaxing the multi-information, $I$, is equivalent to
maximising the joint entropy, $\jointShannonEntropy$, because $\sum_i
\marginalShannonEntropy_i$ is held fixed by information isolation.

The game is played in the exponential family and the gradient flow
occurs in natural parameter space. In natural parameter space, we
perform steepest gradient ascent in the direction dictated by the
Riemannian metric. This ensures that the flow direction is
axiomatically distinguished. Note that in the classical game
\citep{Lawrence-inaccessible25} the steepest entropy ascent was
performed in the Euclidean metric.

The joint entropy of the exponential family can be written as
$$
\jointShannonEntropy = - \naturalParameters^\top
\sufficientStatisticsExpectation +
\cumulantGeneratingFunction(\naturalParameters),
$$
where $\naturalParameters$ are the natural parameters,
$\sufficientStatisticsExpectation$ is the expectation of the
sufficient statistics, $\sufficientStatistics$, and
$\cumulantGeneratingFunction(\cdot)$ is the cumulant generating
function so that $\nabla
\cumulantGeneratingFunction(\naturalParameters) =
\sufficientStatisticsExpectation$.

To find the start of the game we reverse the dynamics and see that the
game starts with the largest possible multi-information, $I$. Playing
the dynamics forward we descend multi-information gradient under the
information isolation constraint. We index the values of the natural
parameters along the flow, $\naturalParameters(\gameTime)$, by
$\gameTime$, an affine parameter that we call \emph{game time}. We
represent the flow start by the `origin', $\gameTime_o$, and the
finish by the `end', $\gameTime_e$.

The end is easily described. The maximum possible value for
$\jointShannonEntropy$ occurs when $I$ is zero (i.e.\ no correlation
between variables) and $\jointShannonEntropy = C$. This is when all
variables are \emph{independent} and \emph{equiprobable}, captured by
a natural parameter vector $\naturalParameters=\mathbf{0}$ and
$\cumulantGeneratingFunction(\naturalParameters) = C$.

The origin is more complicated. Characterising it reveals two
paradoxes, each requiring a conceptual resolution. The first concerns
the existence of the origin state itself; the second concerns the
definition of entropy gradients at that point. We address them in
sequence.

The entropy conservation constraint implies that at least some of the
marginal entropies have to be positive (summing to $C$), but the
structure of Shannon entropy prevents the joint entropy,
$\jointShannonEntropy$, from being zero when the marginal entropies
have a positive sum.

Consider a simple two variable system. The origin constraint demands
non-zero marginal entropy but zero joint entropy. This configuration
is impossible under classical Shannon entropy.

Recall that the multi-information is $I_{1,2} =
\marginalShannonEntropy_1 + \marginalShannonEntropy_2 -
\jointShannonEntropy_{1,2}$. By the chain rule, $I_{1,2} =
\marginalShannonEntropy_1 - \marginalShannonEntropy_{1\mid 2}$, where
$\marginalShannonEntropy_{1\mid 2} = \jointShannonEntropy_{1,2} -
\marginalShannonEntropy_2$ is the conditional entropy. But for Shannon
entropy $\marginalShannonEntropy_{1\mid 2}\geq 0$ always, because
conditional entropy measures residual uncertainty after conditioning,
which cannot be negative. This non-negativity implies $I_{1,2} \leq
\marginalShannonEntropy_1$, and by symmetry, $I_{1,2}\leq
\min(\marginalShannonEntropy_1, \marginalShannonEntropy_2)$. But if
$\jointShannonEntropy_{1,2}=0$, then $I_{1,2} =
\marginalShannonEntropy_1 + \marginalShannonEntropy_2$, which violates
the bound unless both $\marginalShannonEntropy_1 =
\marginalShannonEntropy_2=0$. Therefore zero joint entropy with
non-zero marginals cannot exist classically.

The root cause is that the commuting random variables characterised by
Shannon entropy necessarily obey $\jointShannonEntropy_{X\mid Y} \geq
0$. \citet{Baez-characterisation11}'s three axioms (functoriality,
convex linearity, continuity) uniquely select Shannon entropy, but
Shannon entropy cannot represent the origin.

\subsection{Resolution of the Conflict}
\label{subsec-quantum-resolution}

The resolution requires moving from Shannon entropy to von Neumann
entropy. In von Neumann entropy \citep[][denoted here by
  $\jointVonNeumannEntropy$]{vonNeumann-book32} the conditional
entropy $\jointVonNeumannEntropy_{A|B} = \jointVonNeumannEntropy_{A,B}
- \jointVonNeumannEntropy_{B}$ \emph{can} be negative when subsystems
$A$ and $B$ are `entangled'.

The von Neumann entropy was motivated by a desire to place the
mathematics of quantum mechanics on a more rigorous footing. Here we
are drawn to the representation as a characterisation of
information loss that can tolerate a negative conditional entropy,
$\jointVonNeumannEntropy_{B|A} < 0$ allowing configurations with zero
joint entropy and non-zero marginal entropies,
$\marginalVonNeumannEntropy_A >0$, $\marginalVonNeumannEntropy_B > 0$.

These states have a simple characterisation: the marginal entropy of
any subsystem is bounded by $\marginalVonNeumannEntropy_i \leq \log
d_i$, where $d_i$ is the number of levels in subsystem $i$. Equality
is obtained when the reduced state is \emph{maximally mixed}.  To
obtain a canonical parameter-free notion of the origin, we impose an
additional selection principle: among globally pure states we choose
those that maximise the conserved marginal-entropy sum. Since
$\marginalVonNeumannEntropy_i \le \log d_i$, this is equivalent to
requiring each reduced state to be maximally mixed, which fixes $C$ to
its maximal value $C_{\max}=\sum_i \log d_i$. A globally pure state
saturating these bounds is maximally entangled. With this additional
criterion an axiomatically distinguished origin is selected up to
local-unitary equivalence.

Moving away from Shannon entropy requires us to change the axiomatic
framework underpinning the game. On the classical side, the BFL axioms
characterise entropy as information loss under measure-preserving
functions \citep{Baez-characterisation11}. On the quantum side, the
categorical framework of \citet{Parzygnat-functorial22} plays an
analogous role. It works in the category of finite-dimensional
$C^*$-algebras with state-preserving $*$-homomorphisms as morphisms.
Within this setting, von Neumann entropy arises as the unique
functorial measure of information loss for state-preserving
$*$-homomorphisms, in direct analogy with Shannon entropy for
stochastic maps.

Alternative axiomatisations also characterise von Neumann entropy
uniquely, e.g.\ via unitary invariance, strong subadditivity, and
continuity \citep{Wehrl-entropy78}, but these treat entropy as a state
functional rather than as functorial information loss.  Parzygnat's
work follows in the tradition of the BFL axioms which characterise
entropy as the unique quantity measuring loss of distinguishability
under morphisms giving us Baez--Fritz--Leinster--Parzygnat (BFLP)
categorical frameworks
\citep{Baez-characterisation11,Parzygnat-functorial22}.  For our
purposes, adopting \citet{Parzygnat-functorial22}'s characterisation
of information loss resolves the origin paradox. So we take
\begin{equation}
  \jointVonNeumannEntropy(\rho) = - \tr\left(\rho \log \rho\right),
\end{equation}
where $\rho$ is a \emph{density matrix} and $\log$ is the matrix
logarithm, to be the entropy functional underpinning the updated game's
information loss.

\subsection{An Algebraic Perspective on Probability}
\label{sec-algebraic-perspective}

To place the von Neumann entropy in the context of the inaccessible
game we take an \emph{algebraic perspective} on probability
\citep[see][]{Petz-quantum08,Accardi-quantum82,Holevo-statistical01}.
This changes the usual view where \emph{outcomes} are primitive and
replaces them with \emph{expectations} as the
primitive. Philosophically, this complements the inaccessible game's
framing where the system is information isolated meaning outcomes are
never instantiated.

The density matrix representation can be viewed as an
expectation-based model which doesn't assume the expectations arise
from an underlying sample space. Within the inaccessible game an
expectation-first notion of state is natural. We track
expectations\footnote{Here we use $\langle \cdot \rangle$ to represent
expectation.} $\langle A \rangle$ and their Legendre duals, the
natural parameters $\naturalParameters$. We do not track outcomes.

Classical probability theory assumes observables form a commutative
algebra, equivalent to real-valued functions on a sample space
$X$. Noncommutative probability removes this restriction: sufficient
statistics are replaced by \emph{Hermitian} operators on a Hilbert space
forming a noncommutative $*$-algebra, with expectation values
$\langle A \rangle = \tr(\rho A)$ under density matrices
$\rho$. The noncommutative framework maintains the algebraic
structure without assuming an underlying sample space. Adopting
non-commutative probability for the origin suggests that `outcomes' do
not occur at the pure LME state: there is no joint assignment of
values to all observables compatible with the algebra, only
expectation values.

This perspective aligns with Jaynes' maximum entropy formulation
\citep{Jaynes-information57a,Jaynes-information57b}, where
expectations, not samples, constitute the primitive data. Classical
impossibility results like $\jointShannonEntropy_{X|Y} \geq 0$ depend
on the existence of a joint sample space; relaxing that requirement
allows us to recover the game origin through von Neumann entropy.

\section{Von Neumann Entropy and the Matrix Exponential Family}
\label{sec-matrix-exponential}

The von Neumann entropy pre-dates Shannon's entropy. It was derived as
part of von Neumann's mathematical foundation of quantum mechanics,
\citep{vonNeumann-book32}. Shannon entropy and classical probability
emerge as special cases when the observables of interest mutually
commute and are jointly diagonalisable with the density matrix. In the
resulting common eigenbasis the density matrix is diagonalisable and
can be reduced to a classical probability distribution. This section
reviews standard material on the matrix exponential family and applies
the family to resolve the inaccessible game's origin paradox.

The density matrix is normally represented as,
$$
\rho = \sum_{j=1}^d p_j \ket{\psi_j} \bra{\psi_j}
$$
with $p_j \geq 0$ and $\sum_{j=1}^d p_j = 1$. Here we have introduced
Dirac's bra ket notation,\footnote{Note that for bra and ket $\langle$
and $\rangle$ are each always paired with $\mid$ so there is no
ambiguity with our expectation notation $\langle \cdot \rangle$ which
contains no $\mid$.} where $\ket{\psi_j}$ is a \emph{ket} which
represents the state $\psi_j$ as a column vector and $\bra{\psi_j}$ is
a \emph{bra} which represents the complex conjugate of $\psi_j$ as a
row vector, i.e.\ the Hermitian conjugate. Each $\psi_j$ is a
normalised complex vector associated with a probability $p_j$. The
probability-weighted sum of these forms gives the density matrix and
the entropy is then given by
$$
\jointVonNeumannEntropy = -\tr (\rho \log \rho) = - \sum_{j=1}^d
\densityEigenvalue_j \log \densityEigenvalue_j,
$$
where $\densityEigenvalue_j$ is the $j$th eigenvalue of $\rho$.  For a
fixed set of probabilities $\{p_j\}_{j=1}^d$ and vectors
$\{\ket{\psi_j}\}$, the Shannon entropy $\jointShannonEntropy =
-\sum_j p_j \log p_j$ provides an upper bound on
$\jointVonNeumannEntropy$. Equality holds if and only if the vectors
$\{\ket{\psi_j}\}$ are orthonormal. In bra ket notation we represent
this condition as $\langle \psi_i | \psi_j\rangle=\delta_{ij}$
$\forall i, j$. In this case the eigenvalues of $\rho$ are precisely
$\densityEigenvalue_j = p_j$ restoring commutation and rendering the
two entropies equal. When the vectors overlap entanglement reduces the
von Neumann entropy below its classical upper bound.

For a system of fixed dimension the maximum multi-information is
achieved at a pure LME state: a globally pure state whose reduced
state on each subsystem is maximally mixed. When local Hilbert spaces
have dimension $d_i$, the subsystems are \emph{qudits} each with
dimension $d_i$. The game dynamics then consist of an information
relaxation principle on this substrate of qudits. The information
conservation constraint means that information relaxation is
equivalent to a maximum entropy production principle
\citep{Beretta-quantum85,Ziegler-maximal87,
  Martyushev-maximum06,Beretta-nonlinear09,Beretta-fourth20} under
marginal entropy constraints \citep{Lawrence-inaccessible25}.

The mechanism by which conditional entropies become negative is known
as quantum interference. It renders the algebra noncommutative and
disrupts classical notions of causality, communication, and
conditional independence. As we'll see, these complications leave the
underlying framework of the inaccessible game intact. But the
information relaxation dynamics must be re-derived in the matrix
exponential family coordinates.

\subsection{Matrix Exponential Representation and Constrained Flow}
\label{sec-quantum-flow}

A matrix exponential family can be defined similarly to the classical
exponential family,
\begin{equation}
  \rho(\naturalParameters) = \exp\left( K(\naturalParameters) -
 \cumulantGeneratingFunction(\naturalParameters)\mathbf{I} \right), \qquad
  K(\naturalParameters) = \sum_{a=1}^m \naturalParameter_a F_a,
  \label{eq-quantum-exp-family}
\end{equation}
where $\exp(\cdot)$ is matrix exponentiation and the exponent in
\eqref{eq-quantum-exp-family} is an operator.\footnote{It's more usual
to write $\exp\left( K(\naturalParameters) -
\cumulantGeneratingFunction(\naturalParameters) \right)$ where the
identity operator associated with $
\cumulantGeneratingFunction(\naturalParameters)$ is implied.}  The
sufficient statistics are now given by a Hermitian operator basis
$\{F_a\}_{a=1}^m$ which ensures that the eigenvalues of the exponent
are real and the corresponding density matrix is positive
definite.\footnote{Because all eigenvalues of a Hermitian operator are
real, matrix exponentiation exponentiates the real eigenvalues leading
to positive real eigenvalues and a corresponding positive definite
matrix.} The cumulant generating function has the form
$$
\cumulantGeneratingFunction(\naturalParameters)
= \log \tr\left( \exp\left({K(\naturalParameters)}\right) \right),
$$
where the trace ensures that the
density matrix is normalised.

We use the term matrix exponential family, but this form is more
usually referred to as the \emph{quantum} exponential family. The
cumulant generating function, $\cumulantGeneratingFunction(\cdot)$, is
also more typically known as the log partition function. We prefer
matrix exponential family because the properties we are interested in
do not come from quantisation, but from the noncommutative nature of
the matrix exponential. Specifically, for noncommuting $A$ and $B$,
$\exp(A + B) \neq \exp(A) \exp(B)$. Equality is restored when $A$
and $B$ commute. This has an instrumental effect on computing higher
order cumulants. These are obtained through differentiating the
cumulant generating function which now requires Fr\'{e}chet
derivatives \citep[see e.g.][]{Higham-functions08} due to the presence
of the matrix exponential.

The natural parameters are $\naturalParameters =
(\naturalParameter_a)_a$ and the expectation parameters are
$$
\sufficientStatisticExpectation_a(\naturalParameters) = \tr
\left(\rho(\naturalParameters) F_a\right) = \frac{\partial
  \cumulantGeneratingFunction(\naturalParameters)}{\partial
  \naturalParameter_a}.
$$
When the relevant observable algebra is commutative the matrix
exponential reverts to allowing $\exp(A + B) = \exp(A)\exp(B)$ and the
system reduces to the classical exponential setting. The resulting
diagonal operators correspond to real-valued random variables.

\subsection{Exponential Families and Certainty}

Exponential families cannot represent certainty. For example, from the
classical exponential family consider the Bernoulli distribution,
$p(x) = p^x(1-p)^{(1-x)}$ where $p$ is the probability of $x$ being
1. This takes the exponential family form $p(x) = \exp(\naturalParameter x -
\psi(\naturalParameter))$ with $\naturalParameter =
\log\left(\tfrac{p}{1-p}\right)$ and $\psi = \log (1 +
\exp(\naturalParameter))$. For $x=1$ with certainty ($p = 1$) we would
need $\naturalParameter = \log \tfrac{1}{0} \equiv \infty$ and
similarly for $x=0$ with certainty ($p=0$) we would require
$\naturalParameter = -\infty$.

For the exponential family states of certainty lie on the boundary of
the parametrisation where $\|\naturalParameters\| \rightarrow
\infty$. The family is valid only in the interior.

The same challenge manifests in the noncommuting matrix exponential
family where the boundary also includes pure (non-faithful) states. In
pure states the density matrix is rank deficient. Only in the interior
can we restrict $\rho$ to a full-rank density matrix,
\begin{equation}
  \rho =
  \frac{\exp\left(-K\right)}{\tr\left(\exp\left(-K\right)\right)},
  \label{eq-faithful-state-gibbs}
\end{equation}
then the full-rank (faithful) state is a Gibbs state for its own
modular Hamiltonian,\footnote{For full-rank $\rho$, the modular
Hamiltonian is defined as $K = -\log\rho$ up to an additive constant.}
unique up to the identity shift $K \mapsto K + c\mathbf{I}$.

\subsection{Steepest Ascent}

A density matrix that satisfies maximum von Neumann entropy subject to
linear constraints takes the matrix exponential form
\eqref{eq-quantum-exp-family} via Jaynes's maximum entropy principle
\citep{Jaynes-information57a}.\footnote{Jaynes's principle states that
subject to linear constraints $\langle F_a \rangle =
\sufficientStatisticExpectation_a$, the maximum-entropy density matrix
is $\rho = \exp(K - \cumulantGeneratingFunction)$ where $K = \sum_a
\naturalParameter_a F_a$ encodes the constraints: the natural
parameters are the negative of Jaynes' Lagrange multipliers. The resulting
operator exponential form is the matrix exponential family.} Within
any such exponential family chart, the Fisher information, which
governs the conductance of information flow on the constraint
manifold, is given by the Hessian of the cumulant generating function,
\begin{equation}
G_{ab}(\naturalParameters) = \frac{\partial^2
  \cumulantGeneratingFunction}{\partial\naturalParameter_a
  \partial\naturalParameter_b},
\label{eq-metric-definition}
\end{equation}
which is real, symmetric, and positive definite on full-rank
states. This Hessian is the quantum generalisation of classical Fisher
information \citep[see][for the classical
  result]{Amari-differential82}. In quantum mechanics it is often
known as the Bogoliubov--Kubo--Mori
\citep[BKM,][]{Morozova-markov91,Petz-monotone96} metric. Because it
is the second derivative of the cumulant generating function it is
also the second cumulant of the density matrix. In the literature
these derivatives are known as the \emph{Kubo--Mori cumulants}. Thus
the BKM metric emerges as the natural information geometry
\citep[see][]{Amari-differential82,Amari-information00,Matsuzoe-halfcentury24}.

In the inaccessible game \citep{Lawrence-inaccessible25} the fourth
axiom, i.e.\ information isolation under exchangeability, requires
that marginal entropy sums are conserved, $\sum_i
\marginalVonNeumannEntropy_i = C$. When combined with the information
relaxation principle it leads to constrained steepest entropy
ascent. The pure origin state is justified by reversing these dynamics
and noticing that the minimum joint entropy occurs with maximum
marginal entropy ($C = C_\text{max}$) for the axiomatically
distinguished LME state.

Within the matrix exponential family chart, the von Neumann entropy
can be written as a convex function of the natural parameters whose
gradient is linear in $\naturalParameters$,
$$
\jointVonNeumannEntropy(\naturalParameters) = - \sum_a
\naturalParameter_a \sufficientStatisticExpectation_a +
\cumulantGeneratingFunction(\naturalParameters).
$$
where $\sufficientStatisticExpectation_a = \tr(\rho F_a)$.  In these
coordinates we have $\nabla_{\naturalParameters}
\jointVonNeumannEntropy = -G(\naturalParameters)\naturalParameters$,
where $G$ is the BKM metric. The information relaxation principle
requires a choice of dissipation geometry to make `steepest' well
defined.  The Riemannian metric is axiomatically distinguished and
invariant to coordinate transformation.  The marginal entropy
conservation constraint $\sum_i
\marginalVonNeumannEntropy_i(\naturalParameters) = C$ is enforced via
Lagrange multiplier, yielding the projected flow,
\begin{equation}
\dot{\naturalParameters} = -\Pi_{\text{marg}}(\naturalParameters)
\naturalParameters.
\label{eq-projected-theta-flow}
\end{equation}
Here $\dot{\naturalParameters}$ denotes derivative with respect to the
affine parameter $\gameTime$, and
$\Pi_{\text{marg}}(\naturalParameters)$ denotes the Fisher orthogonal
projector onto the marginal-preserving tangent space. Next we will
show that in the confined $C=C_{\max}$ regime this constrains us to
directions that preserve reduced states to first order,
$\dot{\rho}_i=0$, for every subsystem $i$.

\subsection{Confinement of the Marginal Entropies}
\label{sec:confinement-of-the-marginal-entropies}

Our information-isolation constraint is the conservation of the
marginal-entropy sum
\begin{equation}
  C(\naturalParameters) \coloneq \sum_i
  \marginalVonNeumannEntropy_i(\naturalParameters), \qquad
  C(\naturalParameters(\gameTime)) = C_{\max}.
\label{eq:def-C-marginal-entropy-sum}
\end{equation}
At the axiomatically distinguished LME origin, the single scalar
constraint $C(\naturalParameters) = C_{\max}$ becomes equivalent to a
\emph{system} of constraints. Since each subsystem's marginal entropy
is bounded by $\marginalVonNeumannEntropy_i \leq \log d_i$, saturation
of the sum $C_{\max} = \sum_i \marginalVonNeumannEntropy_i = \sum_i
\log d_i$ can only occur when \emph{each} term individually saturates
its bound. Once they have saturated they become confined: any movement
downwards of subsystem $i$ cannot be compensated by a corresponding
movement upwards of any subsystem from $\bar{i}$ as they are all at
upper bound.

If the marginal-entropy sum is conserved at its maximal value,
$$
C(\rho)=\sum_i \marginalVonNeumannEntropy(\rho_i)=\sum_i \log d_i,
$$
then along any admissible trajectory we have
$$
\rho_i(\gameTime)=\frac{\mathbf{I}_{d_i}}{d_i}\qquad\forall i,\ \forall \gameTime.
$$
Equivalently, the admissible state space at $C=C_{\max}$ is the
fixed-marginals manifold
$$
\mathcal{M}_{\text{marg}} \coloneq
\left\{\rho:\tr_{\bar{i}}\left(\rho\right) =
\frac{\mathbf{I}_{d_i}}{d_i}\ \ \forall i\right\}.
$$
where $\tr_{\bar{i}}(\cdot)$ denotes the partial trace that when
applied leaves us with subsystem $i$. Each term obeys
$\marginalVonNeumannEntropy(\rho_i)\leq \log d_i$, with equality if
and only if $\rho_i=\mathbf{I}_{d_i}/d_i$. If $\sum_i
\marginalVonNeumannEntropy(\rho_i)=\sum_i \log d_i$, then every term
must saturate and hence every $\rho_i$ is maximally mixed.  The LME
origin is not an isolated boundary point, but a distinguished boundary
point of $\mathcal{M}_{\text{marg}}$. The second-order analysis below
diagnoses how the scalar constraint $C=\text{const}$ becomes
insufficient at saturation and how admissibility must be enforced
using the tangent geometry of $\mathcal{M}_{\text{marg}}$.

Our selection criterion fixes the origin as a locally maximally
entangled (LME) pure state: each reduced state is maximally mixed and
each marginal entropy is maximised, $\marginalVonNeumannEntropy_i=\log
d_i$ and $C(\naturalParameters^\ast)$ is saturated,
\begin{equation}
  \mathbf{a}(\naturalParameters^\ast) = \nabla_{\naturalParameters}
  C(\naturalParameters) \big|_{\naturalParameters^\ast} =\mathbf{0}.
\label{eq:a-vanishes-at-saturation}
\end{equation}
The usual first-order tangency condition for a constrained trajectory
$\naturalParameters(\gameTime)$,
\begin{equation}
\frac{\text{d}}{\text{d}\gameTime}C(\naturalParameters(\gameTime))
=\mathbf{a}(\naturalParameters)^\top \dot{\naturalParameters} =0,
\label{eq:first-order-tangency}
\end{equation}
becomes vacuous at $\naturalParameters^\ast$: it is automatically
satisfied for every velocity $\dot{\naturalParameters}$ because the
gradient vanishes. Equivalently, any rank-1 scalar-constraint
projection formula becomes indeterminate and cannot select an
admissible velocity direction at the origin. This degeneracy does
\emph{not} mean the constraint is inactive; it means that the scalar
summary no longer carries enough first-order information to determine
admissibility. At $C=C_{\max}$ the operative constraint is already the
manifold $\mathcal{M}_{\text{marg}}$ of fixed maximally mixed
marginals.

At a saturated extremum, the first nontrivial information about
constraint preservation appears at second order along the
trajectory. We impose
\begin{equation}
\frac{\text{d}^2}{\text{d}\gameTime^2}C(\naturalParameters(\gameTime))
\Big|_{\naturalParameters^\ast}=0.
\label{eq:second-order-invariance}
\end{equation}
Differentiating twice gives the standard decomposition into a
quadratic `velocity' term plus an `acceleration' term so we are left
with
$$
\frac{\text{d}^2}{\text{d}\gameTime^2}C(\naturalParameters(\gameTime))
= \dot{\naturalParameters}^\top \left(\nabla^2
  C(\naturalParameters)\right) \dot{\naturalParameters}
+\mathbf{a}(\naturalParameters)^\top \ddot{\naturalParameters},
$$
and evaluating at a saturated point
\eqref{eq:a-vanishes-at-saturation} eliminates the acceleration term,
\begin{equation}
\dot{\naturalParameters}^\top \left(\nabla^2
  C(\naturalParameters^\ast) \right)\dot{\naturalParameters}=0.
\label{eq:quadratic-velocity-constraint}
\end{equation}
Since $C$ is maximised at $\naturalParameters^\ast$, the Hessian
$\nabla^2 C(\naturalParameters^\ast)$ is negative semidefinite, and
the quadratic constraint \eqref{eq:quadratic-velocity-constraint} is
equivalent to a linear-algebraic restriction on the velocity,
\begin{equation}
\dot{\naturalParameters}\in \mathcal{V}_2(\naturalParameters^\ast)
\coloneq \ker \nabla^2 C(\naturalParameters^\ast) .
\label{eq:V2-kernel-definition}
\end{equation}
This subspace $\mathcal{V}_2$ is the second-order admissible velocity
subspace. It replaces the collapsed first-order hyperplane
$\{\dot{\naturalParameters}:\mathbf{a}^\top\dot{\naturalParameters}=0\}$.  

At the LME origin, the second-order admissibility condition selects
directions that preserve all reduced states to first order. Directions
in $\ker \nabla^2 C(\naturalParameters^\ast)$ are characterised by
inducing $\dot{\rho}_i = 0$ for all subsystems $i$. This connection
arises because at maximal mixing, the first variation of each
$\marginalVonNeumannEntropy_i$ with respect to trace-preserving
perturbations vanishes identically. The second-order invariance of
$C(\naturalParameters^\ast)$ reduces to requiring no first-order
change in any reduced state.

Let $N$ be any matrix whose columns form a basis for
$\mathcal{V}_2(\naturalParameters^\ast)$.  Then the
$G(\naturalParameters^\ast)$-orthogonal projector onto second-order
admissible velocities is
\begin{equation}
\Pi_{\text{marg}}(\naturalParameters^\ast) = N \big(N^\top
G(\naturalParameters^\ast) N \big)^{-1} N^\top
G(\naturalParameters^\ast).
\label{eq:second-order-projector}
\end{equation}

The matrix $N$ has columns forming a basis for the marginal-preserving
subspace in $\naturalParameters$-space. Concretely, the
marginal-preserving condition $\dot{\rho}_i = 0$ for all $i$
translates to a system of linear constraints on
$\dot{\naturalParameters}$. Let $M(\naturalParameters)$ be the
constraint matrix whose rows encode $\nabla_{\naturalParameters}
\text{vec}(\rho_i)$ for each subsystem $i$, where $\text{vec}(\cdot)$
denotes vectorisation of the density matrix. Then the
marginal-preserving tangent space is $\{\dot{\naturalParameters} :
M(\naturalParameters)\dot{\naturalParameters} = 0\}$, and $N$ is any
matrix whose columns span $\ker M(\naturalParameters)$. The projector
\eqref{eq:second-order-projector} is then the $G$-orthogonal projector
onto this kernel. At generic interior points, the standard formula
$\Pi = \mathbf{I} - M^\top(M G^{-1} M^\top)^{-1} M G^{-1}$ applies
when the constraint gradient is nonzero; at saturation the
kernel-based form \eqref{eq:second-order-projector} replaces it.

Near the LME origin, the constrained steepest-entropy-ascent flow is
obtained by projecting the unconstrained natural-parameter contraction
onto $\mathcal{V}_2$,
$$
\dot{\naturalParameters} =
-\Pi_{\text{marg}}(\naturalParameters^\ast) \naturalParameters.
$$

\subsection{Axiomatically Distinguished Trajectory}

Importantly, in the confined regime where $C=C_{\max}$ for all times,
admissible motion is modelled on the fixed-marginals manifold
$\mathcal{M}_{\text{marg}}$. Along $\mathcal{M}_{\text{marg}}$ the
scalar constraint gradient remains zero and the resulting
`origin-distinguished trajectory' lives on this manifold with residual
freedom corresponding to gauge motion along the local-unitary
orbit. At this point we can ensure that this trajectory is
axiomatically distinguished. The information relaxation dynamics from
\citet{Lawrence-inaccessible25} chose the direction of maximum entropy
ascent, though using the Euclidean metric. The axiomatically
distinguished metric of an information geometry is the Riemannian
metric. If we now insist that our steepest direction is measured under
the Riemannian metric, i.e. the Fisher information, then we have both
a starting point, the origin, and an initial direction which together
axiomatically distinguish the trajectory. In the next section we
consider a worked example of this trajectory in a two qutrit system.

\subsection{Worked Example: Two-Qutrit System}
\label{sec:two-qutrit-example}

To make the abstract structure concrete, consider a bipartite system
of two qutrits ($d_1 = d_2 = 3$) with global Hilbert space dimension
$d_1 d_2 = 9$.

The LME origin is any maximally entangled pure state. A canonical
choice is the generalised Bell state
\begin{equation}
|\Phi_3\rangle = \frac{1}{\sqrt{3}}\left(|00\rangle + |11\rangle + |22\rangle\right), \qquad
\rho_{\text{origin}} = |\Phi_3\rangle\langle\Phi_3|.
\end{equation}
This state is globally pure with
$\jointVonNeumannEntropy(\rho_{\text{origin}})=0$. The reduced states
are
$$
\rho_1 = \tr_2(\rho_{\text{origin}}) = \frac{1}{3}\mathbf{I}_3, \qquad
\rho_2 = \tr_1(\rho_{\text{origin}}) = \frac{1}{3}\mathbf{I}_3,
$$
both maximally mixed with marginal entropies
$\marginalVonNeumannEntropy_1 = \marginalVonNeumannEntropy_2 = \log
3$. Hence $C_{\max} = 2\log 3$ and the multi-information is $I=2\log
3$.

At the origin, admissible infinitesimal perturbations must preserve
both marginals to first order: $\dot{\rho}_1=0$ and
$\dot{\rho}_2=0$. Since each $\rho_i=\frac{\mathbf{I}_3}{3}$ is
maximally mixed, any local unitary $U_i\in SU(3)$ acting on subsystem
$i$ alone leaves $\rho_i$ invariant. Therefore the entire
local-unitary orbit
$$
\rho(\alpha) = (U_1(\alpha)\otimes U_2(\alpha)) \rho_{\text{origin}} (U_1(\alpha) \otimes U_2(\alpha))^\dagger,
$$
with $U_i(\alpha)=e^{-\mathrm{i}\alpha K_i}$ and
$K_i\in\mathfrak{su}(3)$ Hermitian acting on qutrit $i$, preserves all
marginals identically and lies in $\mathcal{V}_2$. These are the gauge
directions we later identify in
Section~\ref{sec:gauge-structure-at-origin}. One convenient choice of
basis for $\mathfrak{su}(3)$ is given by the Gell--Mann matrices.

Any direction in $\mathcal{V}_2$ changes the global state while
keeping both $\rho_1$ and $\rho_2$ fixed. For instance, choosing a
local generator $K=\lambda_3^{(1)} \otimes \mathbf{I}^{(2)}$ with
$\lambda_3=\mathrm{diag}(1,-1,0)$ induces
$$
\dot{\rho} = -\mathrm{i}[K, \rho_{\text{origin}}],
$$
which preserves $\rho_1=\frac{\mathbf{I}_3}{3}$ exactly because $K$
is local. More generally, the marginal-preserving subspace consists of
correlation-only perturbations: those whose partial traces vanish,
$\tr_2(\dot{\rho})=0$ and $\tr_1(\dot{\rho})=0$.

As we'd see if we relaxed toward the maximally mixed product state
$\rho_{\text{end}}=\frac{\mathbf{I}_9}{9}$ (maximal joint entropy
$\jointVonNeumannEntropy=\log 9 = 2\log 3$, zero multi-information),
the affine-time trajectory $\naturalParameters(\gameTime)$ diverges as
$\gameTime \to \infty$ when approaching the origin backward.

This example illustrates all key features: the origin as a pure
entangled state, saturation rendering the first-order constraint
vacuous, admissible directions as correlation-only flows preserving
maximally mixed marginals.

We now consider the divergence of $\naturalParameters$ and the
associated affine-time stall as we approach the boundary.  This
boundary effect is the source of the second paradox. The first paradox
was resolved by identifying the origin as a globally pure state. We
additionally applied an axiomatic selection criterion to choose an LME
pure state which distinguished our maximum entropy trajectory.
However, the origin is on the boundary of the full-rank state space
where the natural information-geometric parametrisation becomes
degenerate.

\section{Entropy Time and the Boundary of the Matrix Exponential Family}
\label{sec-entropy-time}

We have identified the game origin with a globally pure state that
maximises the marginal-entropy sum. Unfortunately, the natural parameters
become unbounded and the Riemannian metric (i.e.\ the Fisher
information) degenerates at the origin. The entropy-time
reparametrisation presented here is the second main contribution of this
work: it resolves the geometric singularity at the origin.

The boundary difficulty is resolved not by altering the dynamics, but
by adopting an axiomatically distinguished parametrisation,
$\entropyTime$, of the same constrained flow. We call this
parametrisation \emph{entropy time} because we constrain
$\tfrac{\text{d}\jointVonNeumannEntropy}{\text{d}\entropyTime} = c$
for a fixed constant $c>0$. This ensures that any finite entropy
change corresponds to a finite entropy-time interval.  Throughout,
$\entropyTime$, and the information-geometric expressions used to
define it, are defined on the full-rank interior of state
space. Statements at the pure-state boundary would need to be taken as
limits along full-rank regularisations of trajectories approaching the
origin.

However, even if the natural parametrisation of the boundary state can
only ever be approached and cannot be reached, the origin still
axiomatically distinguishes a trajectory. Whether or not the system
literally starts from the origin, a trajectory associated with the
axiomatically distinguished origin must itself be axiomatically
distinguished.

\subsection{Divergence of the Parameters}

As we approach the origin and $\rho \to \rho_{\text{pure}}$, the
natural parameters diverge: $\|\naturalParameters\| \to \infty$, but
the entropy approaches its global minima of zero.  Since the entropy
$$
\jointVonNeumannEntropy(\naturalParameters) = - \sum_a
\naturalParameter_a \sufficientStatisticExpectation_a +
\cumulantGeneratingFunction(\naturalParameters)
$$
is concave in $\naturalParameters$, and
$\jointVonNeumannEntropy\rightarrow 0$, its global minimum. As we
approach the pure state the concavity of $\jointVonNeumannEntropy$
implies that the gradient must vanish, $\|\nabla
\jointVonNeumannEntropy\| = \|G(\naturalParameters)
\naturalParameters\| \rightarrow 0 $. Since $\naturalParameters$ are
growing in magnitude, this can only happen if the BKM metric develops
a nullspace in the direction of $\naturalParameters$ as the pure state
is approached. The vanishing of the gradient occurs because the BKM
metric degenerates in the direction of diverging parameters.

The game time is an affine parameter, meaning it is only resolved up
to a scale and offset. We can reparametrise by scaling the game time
so that entropy production occurs at a constant rate. We introduce a
new time parameter $\entropyTime$ such that
$$
\frac{\text{d}\jointVonNeumannEntropy}{\text{d}\entropyTime} = c,
$$
along the projected flow~\eqref{eq-projected-theta-flow}.  The choice
of $c$ is a unit convention: different choices correspond to affine
rescalings of $\entropyTime$ and do not alter the integral curves.

Note that this is much stronger than the familiar statement that
`entropy increases with time'. Here the time parameter is defined by
entropy production:\footnote{This interpretation aligns with
\citet{Baez-characterisation11}'s characterisation of entropy as
information loss (and its von Neumann analogue
\citet{Parzygnat-functorial22}); here we use that information-loss
functional to define a preferred clock along the constrained dynamics.
The ``local'' content enters through the additive marginal-entropy
constraint that fixes the information budget subsystem-by-subsystem.}
$\entropyTime$ is the unique reparametrisation, up to affine shift,
for which entropy grows linearly.  In this sense the inaccessible game
construction does not presuppose an external Newtonian time and then
prove entropy monotonicity. Such a time would be expressly forbidden
by information isolation. Instead it singles out a preferred
parametrisation of the constrained information flow in which the
irreversible sector is uniformised. So one unit of $\entropyTime$
equals $c$ nats of entropy production.

The entropy production rate with respect to game time,
$\gameTime$, is
$$
\frac{\text{d}\jointVonNeumannEntropy}{\text{d}\gameTime} =
\naturalParameters^\top G(\naturalParameters)
\Pi_{\text{marg}}(\naturalParameters)\naturalParameters.
$$
Here $\Pi_{\text{marg}}(\naturalParameters)$ denotes the Fisher
orthogonal projector onto the marginal-preserving tangent space, so
that the reduced states are conserved along the projected flow.  Using
the chain rule,
$\tfrac{\text{d}\jointVonNeumannEntropy}{\text{d}\entropyTime} =
\tfrac{\text{d}\jointVonNeumannEntropy}{\text{d}\gameTime} \cdot
\tfrac{\text{d}\gameTime}{\text{d}\entropyTime} = c$, we can solve for
the time dilation factor,
$$
\frac{\text{d}\gameTime}{\text{d}\entropyTime} =
\frac{c}{\naturalParameters^\top G \Pi_{\text{marg}}
  \naturalParameters}.
$$
The vector field in entropy-time coordinates is then
\begin{equation}
\frac{\text{d}\naturalParameters}{\text{d}\entropyTime} =
\frac{\text{d}\naturalParameters}{\text{d}\gameTime}
\frac{\text{d}\gameTime}{\text{d}\entropyTime} =
-\frac{c}{\naturalParameters^\top G(\naturalParameters)
  \Pi_{\text{marg}}(\naturalParameters)
  \naturalParameters} \Pi_{\text{marg}}(\naturalParameters)
  \naturalParameters.
\label{eq-theta-flow-entropy-time}
\end{equation}
The entropy-time reparametrisation \eqref{eq-theta-flow-entropy-time}
is well-defined on any segment of a trajectory for which the entropy
production rate
$\tfrac{\text{d}\jointVonNeumannEntropy}{\text{d}\gameTime}
=\naturalParameters^\top G\Pi_{\text{marg}} \naturalParameters$ is
strictly positive. At points where this rate vanishes
(e.g.\ stationary points of the constrained flow), $\entropyTime$ does
not provide a valid local coordinate.

The entropy time $\entropyTime$ is a nonlinear reparametrisation whose
dilation factor $\tfrac{\text{d}\gameTime}{\text{d}\entropyTime}$ in
\eqref{eq-theta-flow-entropy-time} depends on the local state
$\naturalParameters$. The flow equations
\eqref{eq-theta-flow-entropy-time} describe the same integral curves
as \eqref{eq-projected-theta-flow}, but measured in different units
along the flow.

Next we consider the dynamics of the axiomatically distinguished
trajectory that has emerged from this analysis.

\subsection{Dynamics of the Axiomatically Distinguished Trajectory}
\label{sec:second-order-admissible-directions}

The definition \eqref{eq:V2-kernel-definition} characterises the
admissible velocity directions at the LME origin purely geometrically,
as the nullspace of the second variation of the saturated
marginal-entropy constraint.  We now give this subspace a more
concrete interpretation.

Recall that the marginal-entropy sum $C(\naturalParameters)$ depends
on $\naturalParameters$ only through the reduced density matrices
$\rho_i(\naturalParameters)=\tr_{\bar{i}}\left(\rho(\naturalParameters)\right)$.
At the origin $\naturalParameters^\ast$, each reduced state is
maximally mixed, $\rho_i(\naturalParameters^\ast) =
\tfrac{\mathbf{I}_{d_i}}{d_i}$, and therefore each marginal entropy
$\marginalVonNeumannEntropy_i$ is individually maximised.

At such a maximally mixed state, the first variation of the von
Neumann entropy with respect to trace-preserving perturbations
vanishes identically.  Consequently, the second-order invariance
condition \eqref{eq:second-order-invariance} reduces to the
requirement that admissible velocities induce no first-order change in
any reduced state. In other words, for a velocity
$\dot{\naturalParameters}$ to lie in $\mathcal
V_2(\naturalParameters^\ast)$ it must satisfy
\begin{equation}
\dot{\rho}_i = 0 \qquad \forall i,
\label{eq:marginal-preserving-condition}
\end{equation}
where $\dot{\rho}_i$ denotes the time derivative of the $i$th reduced
density matrix along the trajectory $\naturalParameters(\entropyTime)$.

So the subspace $\mathcal V_2$ consists of directions that change the
global state while preserving all marginals to first order. These
directions correspond to perturbations that act purely on
correlations: they modify joint structure without altering any local
reduced state. This interpretation is independent of coordinate
representation and depends only on the confinement triggered by the
saturation of marginal entropies at the origin.

Unitary evolutions with generators of the local form $K = \sum_i K_i
\otimes \mathbf{I}_{\bar{i}}$ automatically satisfy
\eqref{eq:marginal-preserving-condition} and therefore lie in
$\mathcal V_2(\naturalParameters^\ast)$. Therefore the admissible
directions selected by the second-order constraint include all flows
that preserve reduced states while redistributing correlations. We
return to this connection in Section~\ref{sec:sea-connection}, where
the projected steepest entropy ascent dynamics are decomposed into
reversible and irreversible components.

An important class of such directions arises from the
adjoint action of Hermitian operators. If $\dot{\rho}$ takes the form
\begin{equation}
\dot{\rho} = -\mathrm{i}[K,\rho],
\label{eq:adjoint-action}
\end{equation}
for some Hermitian generator $K$, then $\dot{\rho}$ is automatically
trace-preserving and preserves all reduced states whenever $K$ acts
only through correlations (see also the qutrit example in Section~\ref{sec:two-qutrit-example}). These directions therefore lie in $\mathcal
V_2(\naturalParameters^\ast)$.

Equation \eqref{eq:adjoint-action} is the von Neumann equation in the
Schr\"odinger picture. Its dual, the Heisenberg equation
$\dot{A}=\mathrm{i}[K,A]$, describes the corresponding evolution of
observables.  From the information-geometric perspective adopted here,
these adjoint flows appear as elements of the second-order admissible
velocity subspace singled out by marginal-entropy confinement. The
matrix $N$ entering the projector \eqref{eq:second-order-projector}
has columns whose induced $\dot{\rho}$ lies in the marginal-preserving
subspace.

Importantly, no reference to a Hamiltonian or unitary evolution has
yet been made at the level of assumptions. Adjoint flows of this form
provide a representation of the marginal-preserving directions that
arise from the second-order constraint geometry distinguishing
trajectories initialised at the origin.  When we express the projected
steepest-entropy-ascent dynamics in operator form, this structure
enables a decomposition into reversible and irreversible components,
recovering GENERIC structure.

Returning to our qutrit-pair example we now see that in entropy time
$\entropyTime$ defined by
$\tfrac{\text{d}\jointVonNeumannEntropy}{\text{d}\entropyTime} = c$
for a fixed constant $c>0$, the entropy increases linearly from $0$ to
$2\log 3$ over the finite interval $\entropyTime\in\left[0,
  \tfrac{2\log 3}{c}\right]$ rather than the infinite interval implied
by the game time trajectory, $\gameTime$.

\subsubsection{GENERIC Structure of the Distinguished Trajectory}
\label{sec:gauge-structure-at-origin}

From the information-geometric perspective, the saturated origin
renders local unitary orbits physically indistinguishable across the
entire trajectory. Natural-parameter velocities that differ by
infinitesimal $\mathfrak{su}(d_i)$ actions project to the same
admissible direction once the second-order constraint is
enforced. This redundancy has the structure of a gauge invariance.

The constrained dynamics on the matrix exponential family can be
decomposed into a symmetric steepest entropy ascent sector and an
antisymmetric commutator sector.

The projected dissipative flow
$-\Pi_{\text{marg}}(\naturalParameters)\naturalParameters$ generates
steepest entropy ascent within the marginal-preserving subspace. The
constraint manifold also supports reversible motions that conserve
both von Neumann entropy and all marginal states. In operator form,
these directions are represented by commutator flows $\dot{\rho} =
-\mathrm{i}[K_{\text{local}}, \rho]$ with local generators
$K_{\text{local}} = \sum_i K_i \otimes \mathbf{I}_{\bar{i}}$. The
constrained dynamics then admit a GENERIC-compatible decomposition
combining the dissipative SEA sector with a reversible component from
this gauge orbit.

We write the GENERIC/metriplectic structure using the projector
$\Pi_{\text{marg}}(\naturalParameters)$ onto the marginal-preserving
tangent space in the Fisher geometry. In natural parameter coordinates
$\naturalParameters$, the entropy-time flow takes the form
\begin{equation}
\frac{\text{d}\naturalParameters}{\text{d}\entropyTime} =
\frac{c}{\naturalParameters^\top
    G(\naturalParameters) \Pi_{\text{marg}}(\naturalParameters)
    \naturalParameters}
  \left(\underbrace{-\Pi_{\text{marg}}(\naturalParameters)\naturalParameters}_{\text{dissipative
    flow}} +
\underbrace{\text{ad}_{\xi} \naturalParameters}_{\text{reversible
    flow}}\right).
\label{eq:generic-natural-params}
\end{equation}
Here $\text{ad}_{\xi}$ denotes the adjoint action in
$\naturalParameters$-space corresponding to the commutator $\dot{\rho}
= -\mathrm{i}[\xi,\rho]$ in density-matrix space, where $\xi$ is a
Hermitian generator tangent to the marginal-preserving manifold.

The dissipative flow realises steepest entropy ascent with respect to
the BKM metric restricted to marginal-preserving directions. Along the
origin-distinguished trajectory dissipation is confined to the
correlation sector. It can reduce multi-information while leaving
every marginal maximally mixed.  The reversible flow conserves von
Neumann entropy and is represented by motions tangent to
$\mathcal{M}_{\text{marg}}$; in the simplest case this is the
local-unitary (gauge) orbit generated by a local Hamiltonian
\begin{equation}
  \localHamiltonian(\entropyTime)=\sum_i
  \localHamiltonian_i(\entropyTime)\otimes \mathbf{I}_{\bar{i}},
\end{equation}
for which $\tr_{\bar{i}}\left([ \localHamiltonian(\entropyTime),
  \rho(\entropyTime)]\right) = 0$ holds identically for all subsystems
$i$.

This structure is sometimes called metriplectic or referred to as the
GENERIC (General Equation for Non-Equilibrium Reversible-Irreversible
Coupling) decomposition \citep{Grmela-dynamics97,Ottinger-beyond05}.
It combines a Riemannian metric (governing irreversibility) with a
Poisson bracket (governing reversibility).

For a velocity $v=\dot{\naturalParameters}$, the rate at which the
trajectory attempts to violate the constraint is controlled by the
quadratic form $v^\top
\left(\nabla^2C(\naturalParameters^\ast)\right)v$.  The natural way to
compare constraint curvature against admissible motion is via a
Rayleigh quotient: the denominator $v^\top G v$ measures infinitesimal
distinguishability in the Fisher (BKM) geometry, while the numerator
$v^\top (-\nabla^2 C) v$ measures second-order constraint
violation. Their ratio yields a coordinate-free stiffness
spectrum. Since admissible velocities are measured in the information
geometry induced by the Fisher metric $G$, the intrinsic,
coordinate-invariant measure of second-order constraint violation is
the generalised Rayleigh quotient
\begin{equation}
\kappa(v) = \frac{v^\top
  \left(-\nabla^2C(\naturalParameters^\ast)\right)v}{v^\top
  G(\naturalParameters^\ast)v}.
\label{eq:stiffness-rayleigh}
\end{equation}
Directions with $\kappa(v)=0$ lie in the second-order admissible
subspace $\mathcal V_2(\naturalParameters^\ast)$, while directions
with larger values of $\kappa(v)$ are increasingly confined by the
marginal-entropy constraint. This Rayleigh-quotient characterisation
of constrained flows is standard in information-geometric optimisation
\citep{Amari-differential82,Amari-information00} and appears in
analyses of model parameter spaces and effective theory reductions.
The eigenstructure of the generalised problem $-\nabla^2 C v = \lambda
G v$ provides a canonical decomposition of the tangent space into soft
(constraint-flat) and stiff (constraint-confined) modes without
introducing additional structure beyond the Fisher metric and the
constraint Hessian.

\section{Marginal Entropy as Modular Energy: Connection to Jaynes and Beretta}
\label{sec-classical-vs-quantum-energy}

In this section we look at how our marginal-entropy constraint relates
to the energy constraints used in traditional thermodynamics, such as
Jaynes maximum entropy and Beretta's steepest entropy ascent
(SEAQT). We show that marginal entropy is structurally equivalent to
an expectation constraint on state-dependent site-wise modular
Hamiltonians.\footnote{We previously introduced the modular
Hamiltonian, $K:=-\log\rho$. Here we apply the same construction at
the level of individual subsystems $i$, and then consider the sum
$\sum_i \langle K_i\rangle$ as the conserved quantity.} The
GENERIC/SEAQT frameworks are well-established; what is new here is
their derivation from the inaccessible game's marginal-entropy
conservation axiom.

Information isolation is imposed exactly at the microscopic level. In
the confined $C=C_{\max}$ regime this fixes $\rho_i\equiv
\mathbf{I}_{d_i}/d_i$ and hence $K_i$ is proportional to the
identity. Any appearance of nontrivial Gibbs parameters (i.e.\ an
\emph{effective} $\beta$) can arise only after coarse graining or
restriction to a reduced observable algebra; we will return to this
emergence mechanism in future work.

For each subsystem marginal $\rho_i$, define the local modular
Hamiltonian
\begin{equation}
K_i := -\log \rho_i.
\label{eq:modular-hamiltonian-def}
\end{equation}
Because $K_i$ is a functional calculus of $\rho_i$, it is diagonal in
the same eigenbasis as $\rho_i$ (with eigenvalues $-\log\lambda$ when
$\rho_i$ has eigenvalues $\lambda$). This makes the marginal entropy
an `energy-like' expectation without introducing any additional
operator beyond the state itself. For this case the marginal entropy
is exactly the expectation of this operator with respect to the
marginal state,
\begin{equation}
\marginalVonNeumannEntropy(\rho_i) = -\tr(\rho_i\log\rho_i) = \tr(\rho_i
K_i) = \langle K_i \rangle_{\rho_i}.
\end{equation}
This identity is immediate from the definition of von Neumann
entropy. The reason to introduce it here is to expose the conserved
sum $\sum_i \langle K_i\rangle_{\rho_i}$ as the `energy-like' quantity
implied by information isolation.  Therefore, our constraint
\begin{equation}
\sum_i \marginalVonNeumannEntropy(\rho_i) = C_\text{max}
\end{equation}
is equivalent to constraining the sum of expectations of local modular
Hamiltonians,
\begin{equation}
\sum_i \langle K_i \rangle_{\rho_i} = C_\text{max}.
\label{eq:modular-constraint}
\end{equation}
This definitional rewrite exposes the local additive form of the
conserved quantity: the information-isolation axiom is formulated as a
constraint on marginal entropies under exchangeability, forcing the
conserved `budget' to decompose as a sum over subsystems. In this
sense, the locality of the emergent energy-like conservation law
originates in how information isolation decomposes. Contrast this with
a traditional Hamiltonian that emerges from spatial locality of
interactions.

The identification of modular Hamiltonians with energy-like generators
invites comparison with the thermal time hypothesis
\citep{Connes-vonneumann94,Rovelli-forget11} where a state $\rho$
induces a preferred flow generated by its modular Hamiltonian
$K=-\log\rho$, interpreted as a notion of physical time. Our
construction is closely related in spirit, but differs in a crucial
respect. In the thermal time hypothesis, modular flow is taken to
define time whenever a nontrivial modular generator exists. In
contrast, we show that under information isolation with saturated
marginal entropies, the local modular Hamiltonians reduce to multiples
of the identity and therefore generate \emph{no} physical flow. In
this regime, modular time collapses and no reversible notion of time
exists. Physical time is instead supplied uniquely by entropy
production along the constrained information-relaxation
dynamics. Therefore in the inaccessible game there are conditions
under which modular time is absent. Nontrivial thermal time can only
re-emerge in effective coarse-grained descriptions where approximate
Gibbs structure appears.

\subsection{Energy Constraints in Metastable Regions}

If such coarse-grained meta-stable regions exist, then within the
meta-stable region we obtain a structural identity to Jaynes's
approach of constraining expectations $\langle H\rangle$ in maximum
entropy, or Beretta's approach of conserving mean energy during
steepest entropy ascent. The key difference is that our energy
operators $K_i$ are \emph{state-dependent}, i.e.\ they change as
$\naturalParameters$ evolves, whereas Jaynes and Beretta assume a
fixed external Hamiltonian.

Any bridge back to the standard fixed energy constraint picture must
become a statement about temporal coarse graining. After coarse
graining reduced states may be well-approximated by a Gibbs family
generated by an effectively parameter-invariant operator. Concretely,
if in an effective description for each site $i$ there exists a
Hermitian local generator $\localHamiltonian_i$ and a scalar $\beta$
such that
$$
  \rho_i \approx \frac{1}{Z_i(\beta)}\exp\left(-\beta \localHamiltonian_i\right) ,
  \qquad Z_i(\beta)= \tr\left(\exp\left(-\beta \localHamiltonian_i\right)\right),
$$
then the site-wise modular Hamiltonian becomes approximately linear in
the thermodynamic parameter,
$$
K_i = -\log\rho_i \approx \beta \localHamiltonian_i + \log 
Z_i(\beta) \mathbf{I}.
$$
We refer to this coarse-grained phenomenon as effective Gibbs-lock. It
occurs when in an effective description we can fix a Hermitian
operator $\localHamiltonian_i$ on subsystem $i$. A canonical one
parameter family has the standard identity
$$
  \frac{\text{d}}{\text{d}\beta} \marginalVonNeumannEntropy(\beta)
  = - \beta \text{var}(\localHamiltonian_i),
$$
where the variance here is the second Kubo-Mori cumulant and is
strictly positive. For nontrivial $\localHamiltonian_i$ the entropy is
strictly monotone in $\beta$ on any interval where $\beta \neq
0$. Hence $\marginalVonNeumannEntropy(\rho_i(t))=\text{const}$ implies
$\beta(t)=\text{const}$ whenever $\rho_i(t)$ stays within the Gibbs
family.

It is useful to separate three distinct senses in which ``Gibbs''
language appears in this paper.
\begin{enumerate}[label=(\roman*)]
  \item \label{list-modular-gibbs} \emph{Modular Gibbs (exact,
  identity)}: every full-rank marginal admits the representation
    $\rho_i = e^{-K_i}/\tr(e^{-K_i})$ with $K_i=-\log\rho_i$.
  \item \label{list-trivial-gibbs} \emph{Trivial Gibbs-lock at maximal
  mixing (exact but degenerate)}: if $\rho_i=\mathbf{I}_{d_i}/d_i$
    then $\rho_i$ is Gibbs for \emph{any} Hermitian
    $\localHamiltonian_i$ at $\beta=0$ (up to an additive constant).
  \item \label{list-coarse-grained-gibbs} \emph{Nontrivial effective
  Gibbs-lock (coarse-grained)}: in an effective description along a
    trajectory one may have $K_i(\gameTime)\approx \beta
    \localHamiltonian_i + c_i(\gameTime)\mathbf{I}$ with
    parameter-invariant generators $\localHamiltonian_i$ and
    approximately constant inverse temperature $\beta$.
\end{enumerate}
The origin-distinguished ($C=C_{\max}$) trajectories exhibit
\ref{list-trivial-gibbs} exactly and globally;
\ref{list-coarse-grained-gibbs} can only arise after temporal coarse
graining.\footnote{For \ref{list-coarse-grained-gibbs} to apply, the
subsystems $i$ cannot remain marginally confined at maximal
mixing. Coarse graining could either relax marginal confinement or
lead to agglomeration of subsystems to form effective local
Hamiltonians.}

Effective Gibbs-lock is intentionally conditional: it is a statement
about an effective description along a fixed Gibbs manifold
parameterised by a parameter-invariant generator
$\localHamiltonian_i$.  For the exact constrained information
dynamics, this hypothesis does not hold at $C=C_{\max}$ (where
$\rho_i\equiv\mathbf{I}/d_i$) and is not generically preserved
elsewhere except in frozen dynamics. Thus effective Gibbs-lock can
only be an emergent, coarse-grained regime. The density matrix
$\rho_i(t)$ may remain close to a Gibbs family in an effective
description and $\beta(t)$ may be approximately constant over long
timeframes, but not as an exact identity in the nontrivial sense
\ref{list-coarse-grained-gibbs} above. This emergence of approximate
thermodynamic structure from exact information dynamics is the
mechanism by which irreversibility enters effective thermodynamic
descriptions. We defer detailed analysis of this coarse-graining
mechanism to future work.

The entropy-time parametrisation $\entropyTime$, defined by
$\tfrac{\text{d}\jointVonNeumannEntropy}{\text{d}\entropyTime}=c$ for
a fixed constant $c>0$, fixes an operationally distinguished clock for
the constrained information flow independently of any Hamiltonian
notion of energy.

In the exact confined regime at $C=C_{\max}$, the reversible
(commutator) sector is gauge-like and admits no intrinsic calibration
relative to entropy time; all local modular Hamiltonians reduce to
multiples of the identity, so no nontrivial local generator
exists. Entropy time therefore governs the full physically
distinguishable evolution along this trajectory.

In an effective Gibbs-locked regime, the inverse temperature $\beta$
can calibrate the otherwise free scale of reversible evolution: it
converts an effective entropy-based generator into an effective
physical Hamiltonian $\localHamiltonian_i$ with a definite rate
relative to entropy time. The temperature does not define the clock,
but it can fix the numerical relation between entropy time and
Hamiltonian time in such coarse-grained descriptions. However, in
these systems reversibility can only be approximate and entropy must
be produced. This allows classical thermodynamic time scales to emerge
when this calibration becomes stable, allowing familiar
energy--temperature structures to arise as effective descriptions
within the game's underlying entropy-timed information dynamics.

In the classical inaccessible game \citep[Section
  5]{Lawrence-inaccessible25}, the link between marginal-entropy
conservation and energy conservation emerges asymptotically in the
thermodynamic limit. Classical sufficient statistics
$\sufficientStatistics(\variables)$ define energy functionals
$E(\variables) =
-\boldsymbol{\alpha}^\top\sufficientStatistics(\variables)$, and along
macroscopic order-parameter directions (where multi-information
gradient scales intensively), the constraint gradient becomes
asymptotically parallel to the energy gradient: $\nabla(\sum_i
\marginalShannonEntropy_i) \parallel \nabla E$. This ensures the
degeneracy condition $S\nabla E = 0$ holds in the thermodynamic
limit. This classical alignment is conditional: it requires
macroscopic (large-$n$) systems with order parameters in
near-equilibrium configurations where $-\log p \approx \beta E + \log
Z$, and finite correlation length (clustering) to ensure connected
cumulants dominate and the Fisher metric remains well-defined.

In the quantum setting, the parallel via modular Hamiltonians is not
asymptotic. It is immediate and structural. The marginal-entropy
constraint \eqref{eq:modular-constraint} is already an expectation
constraint on Hermitian operators, so it fits directly into the
Jaynes/Beretta framework without requiring thermodynamic limits. The
reversible sector generates evolution that conserves this
modular-energy sum, which takes the form of unitary evolution. The key
structural distinction is that we do not postulate a fixed energy
observable; instead, the `energy' emerges as the state-dependent
logarithm of the marginal density matrices. In effective Gibbs-locked
regimes, this energy-like quantity could behave approximately like a
scaled fixed Hamiltonian. This makes the constraint self-consistent:
marginal entropies are both the information being protected and, via
the modular identification, the energy-like quantities being
conserved.

\subsection{Steepest entropy ascent and the symmetric sector}
\label{sec:sea-connection}

The symmetric component of the constrained von Neumann entropy flow
derived above is a steepest-entropy-ascent construction in the BKM
geometry. This connects the inaccessible-game dissipative sector to
Beretta's nonlinear steepest entropy ascent dynamics (SEAQT).

In particular, making coefficient-level contact between the linearised
projected flow and standard master-equation forms requires an explicit
pushforward from $\naturalParameters$-space to $\rho$-space for a
chosen operator basis, such as the Gell--Mann basis in our two-qutrit
example, Section~\ref{sec:two-qutrit-example}. We leave that detailed
matching for future work.

At the level of mechanism, the symmetric sector in this paper is
simply steepest entropy ascent in the BKM geometry with the
marginal-entropy constraint providing the conserved energy-like
quantity via modular Hamiltonians
(Section~\ref{sec-classical-vs-quantum-energy}). This is the SEAQT
construction in Beretta's sense
\citep{Beretta-quantum85,Beretta-nonlinear09,Beretta-fourth20}; see
also \citet{Beretta-generic14} for the SEA--GENERIC correspondence. We
have identified the antisymmetric component $A$ as standard unitary
von Neumann evolution under the commutator/Lie--Poisson structure,
completing the GENERIC structure. In any effective Gibbs-locked
regimes (from e.g.\ coarse graining effects), the temporal evolution
of the antisymmetric portion can align with entropy time through an
effective local inverse temperature $\beta$.

\section{Discussion and Outlook}

We have explored the origin of the inaccessible game where
marginal-entropy conservation constrains steepest entropy ascent in
the Riemannian geometry. Seeking a non-degenerate origin forces a
noncommutative setting. Replacing Shannon entropy with von Neumann
entropy resolves the Shannon-origin obstruction. Selecting the maximal
constraint ($C=C_\text{max}$) imposes an locally maximally entangled
(LME) origin: a globally pure state with maximally mixed marginals
defined up to the local-unitary orbit. At this saturated point the
first-order scalar constraint degenerates and a second-order lock
emerges: admissible directions are determined by tangency to the
fixed-marginals manifold, so dissipation persists but is confined to
marginal-preserving correlational directions. In the confined regime
$C=C_{\max}$ the dynamics are modelled on the fixed
maximally-mixed-marginals manifold $\mathcal{M}_{\text{marg}}$ where
termwise saturation of the marginal entropy bounds holds. The scalar
constraint then acts through second-order geometry. Entropy production
determines the dynamics only up to local-unitary gauge: motion along
the local-unitary orbit is invisible to entropy time and remains a
choice of reversible generator. This freedom is naturally represented
by an antisymmetric commutator/Lie--Poisson sector alongside a
symmetric SEA sector, yielding a GENERIC-compatible
reversible--irreversible decomposition. The entropy-time
reparametrisation then provides an operationally distinguished clock
for the trajectory.

Entropy time $\entropyTime$ is defined by fixing the rate of entropy
production,
$\tfrac{\text{d}\jointVonNeumannEntropy}{\text{d}\entropyTime} = c$
for a constant $c>0$. The choice of $c$ is, at the level of the
axioms, a unit convention (an affine rescaling of the trajectory-wise
clock).  In effective descriptions obtained by temporal coarse
graining, one might ask whether a model-dependent minimal temporal
resolution exists below which entropy-producing dynamics become
unresolvable. Such questions lie outside the scope of this paper, but
will be addressed in future work on coarse-grained effective theories.

We have shown that the origin of the inaccessible game requires a
quantum framework: von Neumann entropy admits an LME origin (globally
pure yet with maximally mixed marginals), forbidden classically by
$\jointShannonEntropy(X|Y)\geq 0$. Crucially, selecting the origin
fixes an equivalence class, not a unique state.  At saturation the
first nontrivial constraint structure is second order, and the
Fisher-rescaled curvature spectrum $\kappa(v)$ in
\eqref{eq:stiffness-rayleigh} canonically separates stiff and soft
directions.

The classical inaccessible game \citep{Lawrence-inaccessible25}
motivated a connection between marginal-entropy conservation and
energy conservation in appropriate thermodynamic regimes. In the
quantum setting this bridge is immediate and structural via modular
Hamiltonians.

This makes the marginal-entropy constraint resemble what we've
identified as an energy-like constraint, but with an important
distinction. In the game $K_i$ is state-dependent. In effective
coarse-grained descriptions where reduced states become approximately
Gibbs, $\rho_i \approx \tfrac{e^{-\beta \localHamiltonian_i}}{Z_i}$, the
local modular Hamiltonian would approach $K_i \approx \beta
\localHamiltonian_i + \log Z_i$, so conserving marginal entropies becomes
closely related to conserving (appropriately scaled) local energies up
to an additive constant.

In contrast, the modular Gibbs identity $K_i=-\log\rho_i$ holds
exactly for every full-rank marginal, and at $C=C_{\max}$ the trivial
maximal-mixing `Gibbs-lock' holds exactly and globally (each
$\rho_i=\tfrac{\mathbf{I}_{d_i}}{d_i}$, equivalently $\beta=0$ for any
fixed local generator).

An `approximately Gibbs' regime for the marginals is not implied by
the exact microscopic inaccessible-game axioms; it would have to be an
emergent property of coarse-grained effective descriptions. When such
an effective regime exists, a pronounced separation in the stiffness
spectrum $\kappa(v)$ \eqref{eq:stiffness-rayleigh} would stabilise
effective parameters along soft directions while stiff directions are
slaved to the constraint geometry, making the modular generators $K_i$
behave approximately as scaled fixed Hamiltonians in the reduced
description.

The GENERIC structure we derived is not novel in itself. Quantum
GENERIC and metriplectic formulations have been developed by
\citet{Ottinger-thermodynamic10}, i.e.\ the thermodynamic quantum
master equation using energy constraints and
\citet{Beretta-quantum85,Beretta-fourth20}, i.e.\ steepest entropy
ascent quantum thermodynamics (SEAQT), and in metriplectic/Lindblad
frameworks \citep{Morrison-bracket84,Grmela-dynamics97}. But the path
by which we arrive at this structure is novel. In effective
Gibbs-locked regimes the latent antisymmetric time parameter could
become exposed and calibrated by an effective inverse temperature.

An interesting avenue for further work notes that the Rayleigh
coefficient form for the stiffness spectrum $\kappa(v)$ suggests a
distinguished split between soft and stiff directions for entropy
production. Better understanding of these regimes could provide
reduced descriptions of the game that rely only on the classical
exponential family, thereby reintroducing classical notions of
probability and causality to the game's evolution.

\bibliography{the-inaccessible-game-origin}

\end{document}